\documentclass[twocolumn,prb,showpacs,amsmath,amssymb]{revtex4}

\usepackage{graphicx}

\usepackage{dcolumn}

\usepackage{bm}

\begin{document}


\title{Static and Dynamic Spectroscopy of (Al,Ga)As/GaAs Microdisk Lasers with Interface Fluctuation Quantum Dots}

\author{W.H. Wang$^1$}


\author{S. Ghosh$^2$}

\author{F. M. Mendoza$^2$}

\author{X. Li$^1$}

\author{D. D. Awschalom$^2$}

\author{N. Samarth$^{1,2}$}
\email{nsamarth@psu.edu}

\affiliation{$^1$ Materials Research Institute, Penn State University, University Park PA 16802\\
$^2$Center for Spintronics and Quantum Computation, University of
California, Santa Barbara CA 93106}


\date{\today}

\begin{abstract}

We have studied the steady state and dynamic optical properties of
semiconductor microdisk lasers whose active region contains
interface fluctuation quantum dots in GaAs/(Ga,Al)As quantum
wells. Steady-state measurements of the stimulated emission via
whispering gallery modes yield a quality factor $Q \sim 5600$ and
a coupling constant $\beta \sim 0.09$. The broad gain spectrum
produces mode hopping between spectrally adjacent whispering
gallery modes as a function of temperature and excitation power.
Time- and energy-resolved photoluminescence measurements show that
the emission rise and decay rates increase significantly with
excitation power. Marked differences are observed between the
radiative decay rates in processed and unprocessed samples.
\end{abstract}

\maketitle

Current interest in quantum information processing has motivated several fundamental studies of the coupling between zero dimensional (0D) electronic states and confined photon modes in semiconductor microdisks.\cite{gayral_APL_1999,imamoglu_PRL_1999,michler_APL_2000,michler_Science_2000,cao_APL_2000,young_PRB_2002,gayral_APL_2001} These experiments have focused on high finesse microdisks where the 0D states of self-assembled InAs quantum dots (QDs) couple with the whispering gallery modes (WGMs) of a cylindrical microcavity, and are characterized by a large coupling constant ($\beta \sim 0.1$) between the spontaneous emission and the cavity modes. Because the QD emission is characterized by a narrow homogeneous linewidth while the gain spectrum is characterized by a much larger inhomogeneous broadening,\cite{harris_APL_1998} such microdisks exhibit simultaneous stimulated emission from several whispering gallery cavity modes.\cite{young_PRB_2002} This behavior contrasts with microdisks containing two-dimensional (2D) quantum well (QW) states that typically show stimulated emission from a single cavity mode due to the narrower gain width, but suffer from lower $Q$ (typically between $\sim 3000$ and $\sim 5000$) because of larger absorption in the QW.\cite{mohideen_APL_1994,luo_APL_2000}  Here, we discuss the spontaneous and stimulated emission from microdisks that lie in an intermediate regime, where the active region contains interface fluctuation quantum dots (IFQDs) within narrow GaAs quantum wells.\cite{brunner_PRL_1994,brunner_APL_1994,gammon_APL_1995}   This design drastically reduces the number of modes that lase simultaneously, while allowing for a relatively high $Q$-factor ($Q \sim 5600$) and large coupling constant ($\beta \sim 0.09$). We note that although recent experiments have studied triggered single photon emission from IFQDs incorporated into patterned disks, we are unaware of any demonstrated mode coupling and/or stimulated emission in such structures.\cite{hours_APL_2003}

The microdisks in this study are patterned from GaAs/(Ga,Al)As
heterostructures grown by molecular beam epitaxy (MBE) on (001)
semi-insulating GaAs substrates. The samples consist of a 120 nm
thick GaAs buffer layer, a 500 nm thick Al$_{0.8}$Ga$_{0.2}$As
pedestal layer, and finally a 110 nm thick
Al$_{0.31}$Ga$_{0.69}$As/GaAs heterostructure that forms the
microdisk itself. The microdisk region contains an optically
active region with six 4.2 nm thick GaAs QWs, isolated from each
other by 10 nm thick Al$_{0.31}$Ga$_{0.69}$As barriers. This
optically active region is sandwiched between 38 nm thick
Al$_{0.31}$Ga$_{0.69}$As  layers. At each interface, growth
interruptions of 30 seconds are used to induce IFQDs. As shown in
Fig. 1(a) (top), low temperature micro-photoluminescence
(micro-PL) measurements on such samples reveal sharp spectral
features with a full width at half maximum (FWHM) limited by the
resolution of our spectrometer ($\sim 200 \mu$ eV). We also
fabricated control samples with the optically active region
containing either 10 nm wide QWs or 4.2 nm QWs with no growth
interruptions. In the former, the confined electronic states are
explicitly 2D and do not show any spectral structure in micro-PL
measurements, while in the latter the density and extent of the
lateral 0D confining region is highly reduced, and there are only
weak signs of spectral structure in the micro-PL (Fig. 1 (a)
(bottom)).

Microdisks are processed from the MBE-grown wafers using standard photolithography procedures, followed by a two-step HBr and (NH$_4$)$_2$S wet chemical etch. In the first step, a nonselective HBr-based mesa etch defines the circular portion of the microdisk. The isotropic etching characteristics of HBr reduce the disk diameter by approximately 1 $\mu$m during the etching process. In the second step, the 20-24$\%$ (NH$_4$)$_2$S solution with excess Sulfur (5 mg/ml) etches the Al$_{0.8}$Ga$_{0.2}$As layer with high selectivity, producing microdisks of $\sim$ 4 $\mu \rm{m}$ diameter on narrow pedestals (Fig. 1(b)). The (NH$_4$)$_2$S also passivates the surface states to reduce non-radiative recombination at the microdisk surface. We note that the sole use of (NH$_4$)$_2$S for both selective etching and passivation simplifies the overall processing in comparison with past practices that used a separate selective etch, followed by passivation.\cite{mohideen_APL_1994}  Further, we use a lower temperature ($\sim$50 $^{\circ}$C) for the (NH$_4$)$_2$S solution; this appears to improve the passivation of surface states, resulting in a higher radiative efficiency. Finally, we stabilize the sulfide layers by encapsulating the sample with 40 nm SiN$_x$ deposited using electron cyclotron resonance plasma enhanced chemical vapor deposition.

\begin{figure}[h]
\includegraphics{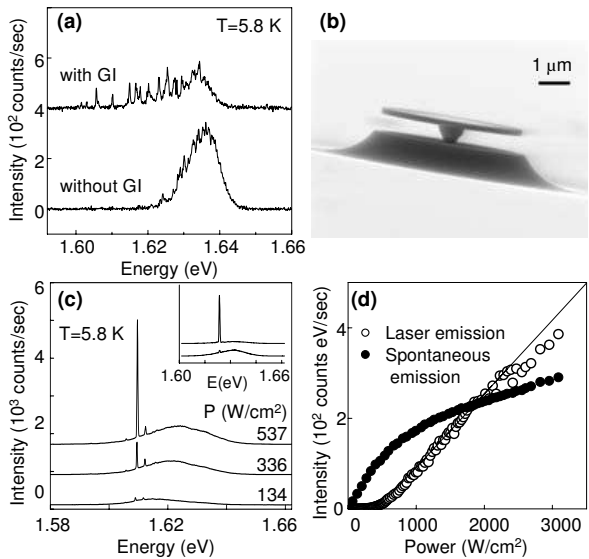}
\caption{(a) Micro-photoluminescence measurements carried out on
unpatterned samples through a 1 $\mu$m diameter gold aperture.
Resolution-limited spectral features characteristic of QD states
(FWHM $ \leq 200 \mu$eV) are only seen in the sample with growth
interruptions. (b) SEM image of microdisk laser structure.  (c)
Excitation power dependence of the emission spectrum of a single
microdisk at $T = 5.8$ K. The inset shows the
polarization-resolved far field emission from the side of a single
microdisk. The top and the bottom spectra are polarization
parallel and perpendicular to the disk plane. (d) Intensity of the
stimulated emission (mode TE$_{35,1}$) and spontaneous emission
vs. pump power. respectively.}
\label{fig1}
\end{figure}

We perform steady-state photoluminescence (PL) measurements using
continuous wave (CW) 568 nm excitation from an Ar/Kr mixed gas
laser. The samples are mounted on a cold finger in a liquid He
continuous flow cryostat, with a temperature sensor co-mounted on
the cold finger close to the sample to monitor heating effects.
Both the optical excitation and collection use an objective lens
(100X, numerical aperture 0.73) with an excitation spot diameter
of 20 $\mu$m; the collected PL is spectrally resolved by a
monochromator (spectral resolution = 180 $\mu$eV) and detected by
a LN2 cooled charge coupled device detector. Polarization-resolved
PL measurements are carried out
 in the far field in a different optical cryostat. Time-resolved PL measurements are made with similar
 excitation/collection optics, but the optical pumping is carried out with a pulsed Ti-Sapphire laser at 740 nm
 (repetition rate of 76 MHz and pulse width of 150 fs).  The PL is again spectrally resolved with a
 monochromator, and
 then temporally resolved using a streak camera,
 which is phase-locked to the pulsed laser in order to synchronize the sweep pulse of the streak tube with the incoming
 signal. The time resolution of the streak camera is 2 ps.

Figure 1(c) shows the CW emission spectra of a single microdisk
laser for three different excitation powers. At the lowest
excitation power, the spectrum is dominated by the spontaneous
emission from the active region, but we already see evidence for
the coupling of the PL emission to a few cavity modes (identified
later as WGMs).
 As the power increases, one of the WGMs lases, with the onset of lasing defined by a clear threshold (Fig. 1(d)).
 The intensity of the integrated PL continues to increase above the lasing threshold; this contrasts with the
 carrier pinning effect observed in conventional semiconductor lasers and is probably due to the non-equilibrium
 heating of the gain medium, as has been reported in other high coupling constant microdisks.
 \cite{henry_IJQE_1982,bjork_APL_1992} The stimulated emission is observed in the same spectral region where the
 IFQD features occur in the micro-PL from unprocessed samples. However, we note that -- unlike the case of microdisks
 containing InAs QDs where the QD emission is well-separated from any wetting layer emission -- it is difficult to
 unambiguously identify IFQD states as the sole source of WGM emission.

Because it is not possible to obtain an exact analytical solution
for the modes of a cylindrical dielectric slab, we use a
well-known approximation that treats the problem as a guided wave
with vacuum wavelength $\lambda$ travelling along the axial
direction ($z$) of a dielectric slab waveguide of refractive index
$n_{\rm{eff}}$, thickness $D$ and radius $R$.\cite{chin_JAP_1994}
The resulting eigenfrequencies are characterized by a set of three
mode numbers $(m, l, p)$ corresponding to the azimuthal, radial
and planar degrees of freedom, respectively. We ignore the planar
mode number $p$ because only the lowest order planar mode ($p =
0$) in the $z$~ direction is allowed by the thin microdisk
($D/\lambda \sim 0.14$).  We estimate the refractive index in the
axial direction ($n_{z}$) from  $\tan \left[ \frac{k_z D}{2}
\right] = \frac{h}{k_z}$, where $k_z$~ is the wave vector in the
$z$~ direction, and ${k_z}^2 + h^2 \equiv k^2 \left[ {n_1}^2 +
{n_2}^2 \right]$, with $n_1$~ and $n_2$~  being the refractive
indices inside and outside the disk structure respectively. The
effective index in the azimuthal direction is then calculated from
$n_{eff}=\alpha \sqrt{{n_1}^2 - {n_z}^2}$,
$\alpha=f-\frac{g}{R/\lambda}$ with parameters $f=0.984$,
$g=0.163$ for $l=1$.\cite{chin_JAP_1994} This results in
$n_{\rm{eff}} \sim 2.3$, compared to the bulk value $n = 3.4$. We
then use this value of $n_{\rm{eff}}$~ to estimate the allowed
eigenfrequencies by solving the two dimensional Helmholtz
equation, whose solution results in Bessel functions of the first
kind.\cite{slusher_APL_1993}  This calculation identifies the
laser emission at 1.61 eV as a WGM with TE polarization and $m =
35, l = 1$. The far field laser emission collected from the side
of the microdisk is strongly polarized in the disk plane (inset to
Fig. 1(c)), confirming the TE nature of the stimulated emission.
We note that selection rules for recombination between conduction
band electrons and heavy holes also produce the same polarization
for spontaneous emission in this geometry, suggesting a possible
reason for the preferred coupling of the emission to a TE rather
than a TM cavity mode.

\begin{figure}[h]
\includegraphics{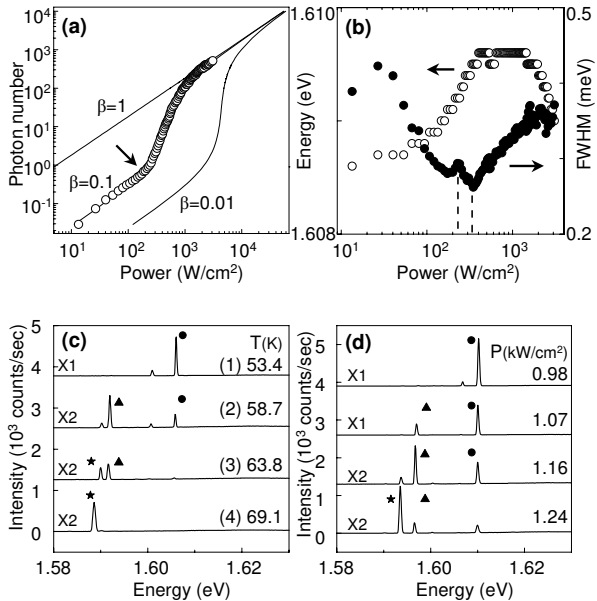}
\caption{(a) Power dependence of lasing intensity plotted on a
log-log scale. The arrow indicates the laser threshold. Solid
curves are based on equation (1) with coupling constant $\beta =
1, 0.1,0.01$. (b) Power dependence of the energy (open circles)
and width (solid circles) of the lasing mode. (c) Temperature
dependence of microdisk emission spectrum, showing mode hopping
between the TE$_{35,1}$ (circle), TE$_{30,2}$ (triangle) and
TE$_{26,3}$ (star) WGMs. (d) Pump power dependence of microdisk
emission spectra at a sample mount temperature of $T = 5$ K. }
\label{fig2}
\end{figure}

Figure 2(a) shows a log-log plot of the the laser emission
intensity in this mode as a function of the excitation intensity
$I$. The laser threshold ($\sim 240$ W/cm$^2$) is indicated by a
kink in this plot, highlighted by the arrow. (As a comparison, we
note that microdisks patterned from the 4.2 nm and 10 nm QW
control samples exhibit somewhat higher thresholds of 410 W/cm$^2$
and 300 W/cm$^2$, respectively.) The laser emission intensity is
proportional to the photon number $p$~ in the mode, which we
define as unity at the threshold.\cite{bjork_PRA_1994} To
determine the cavity coupling constant $\beta$, we fit the data in
Fig. 2(a) using the following equation which relates the photon
number to the excitation intensity:\cite{bjork_IJQE_1991}

\begin{equation}
I (p) = A [\frac{p}{{1 + p}}(1 + \xi )(1 + \beta p) - \xi \beta
p].
\end{equation}

Here, the scale factor $A = \hbar \omega \gamma/\delta \beta$,
where $\omega$~ is the frequency of the mode, $\gamma$~ is the
cavity decay rate, and $\delta$~ is the photon conversion
efficiency. The dimensionless parameter $\xi$~ is defined by $\xi
= N_0 \beta V/\gamma \tau _{\rm{sp}}$, where $N_0$~ is the
transparency carrier concentration of the gain material, $V$~ is
the volume of the active material and $\tau _{\rm{sp}}$ is the
spontaneous emission lifetime. The above equation assumes that
only one mode lases within the gain region, and that non-radiative
recombination is negligible. Since we do not have direct knowledge
of several parameters such as $\gamma$, $N_0$~ and $\delta$, we
treat $\beta$, $\xi$ and $A$ as fitting parameters; the first two
determine the shape of the function $I(p)$, while the latter only
scales its overall magnitude. Our fits yield a coupling constant
$\beta \sim 0.09$~ and $\xi \sim 7$.\cite{note1} For comparison,
we also plot the theoretical curves expected for $\beta  = 1, 0.1,
0.01$. In a conventional laser where $\beta \sim 10^{-3}$, the
photon number typically undergoes a sudden jump at threshold.
Because of the high coupling constant  of the cavity, the output
power in these microdisk lasers increases with a finite slope
around the transition region as shown in Fig. 2(a).

Figure 2 (b) shows the power dependence of the energy and the full
width at half maximum (FWHM) of the TE$_{35, 1}$~ laser mode. For
pump power below 0.4 kW/cm$^2$, the peak energy undergoes a blue
shift with increasing pump power because the refractive index
decreases with increasing carrier density.\cite{mohideen_PRL_1994}
However, the plateau in the peak energy between 0.4 kW/cm$^2$ and
1.4 kW/cm$^2$ indicates that the blue shift is offset by local
heating of the microdisk, eventually resulting in a red shift at
the highest excitation densities.  Below the threshold, we observe a spectral narrowing of the WGM as the excitation power increases. This indicates that the $Q$ is limited by the absorption of the active layer at weak excitation.\cite{gayral_APL_1999,michler_APL_2000} Hence, it is more meaningful to estimate the $Q$~ of the microdisk at the transparency threshold which should lie just below the lasing threshold. At this excitation density, the FWHM is 0.288 meV while the emission wavelength is 1.61 eV, yielding an estimated $Q = \frac{E}{\Delta E} \sim 5600$ for this mode. In comparison, similar estimates for the control samples with 4.2 nm and 10 nm QWs yield $Q \sim 3000$ and $3500$, respectively. Figure 2(b) shows that the FWHM undergoes an anomalous linewidth enhancement in the threshold region. The coupling between refractive index fluctuations and the carrier fluctuations in the gain medium may explain this broadening.\cite{bjork_APL_1992} Above the threshold, the linewidth is inversely proportional to excitation power before the onset of heating, as indicated in the region between the two dashed lines in Fig.2(b). This is described by a modification of the standard Schawlow-Townes formula, wherein the linewidth is multiplied by a factor  $ (1 + \alpha^2)/2$, where $\alpha$ is the gain-refractive index coupling factor.  

The temperature dependence of the emission spectra is measured in
the range $6 \rm{K} \leq T \leq 70 \rm{K}$ with a fixed CW
excitation of 1.8 kW/cm$^2$. Figure 2(c) shows that at high
temperatures (between 50 K and 70 K), the laser emission cascades
toward lower energies via three adjacent WGMs located between 1.61
eV and 1.59 eV (identified as TE$_{35, 1}$, TE$_{30, 2}$ and
TE$_{26, 3}$, respectively). Note that in spectra (1) and (2) of
Fig. 2(c), the emission intensity of the TE$_{35, 1}$~ mode
undergoes an abrupt transition toward the TE$_{30, 2}$~ mode,
without exciting the higher order mode located in between. We
attribute the observed mode switching to a combination of effects:
the temperature dependence of the gain profile, narrowing of the
bandgap and  possibly the thermal excitation of electrons/holes
from deeply confined 0D states to shallower QW states.  We also
observe mode hopping at low temperatures when the excitation power
is large compared to the single-mode operation region discussed
earlier (Fig. 2(d)). Although the power-dependent mode hopping at
low temperatures is also likely due to local heating of the
microdisk (no changes are observed in the temperature sensor), we
note that the transition from the TE$_{35, 1}$~ to the TE$_{30,
2}$ mode is gradual, contrasting with the temperature-dependent
changes shown in Fig. 2(c). We attribute this difference to the
continued occupation of the 0D states at low temperatures as the
pump power increases, as opposed to being depleted to higher
states in fixed power/variable temperature measurements.
Therefore, the TE$_{35, 1}$~ mode does not quench completely and
the high carrier density can sustain multiple-mode operation.

\begin{figure}[h]
\includegraphics{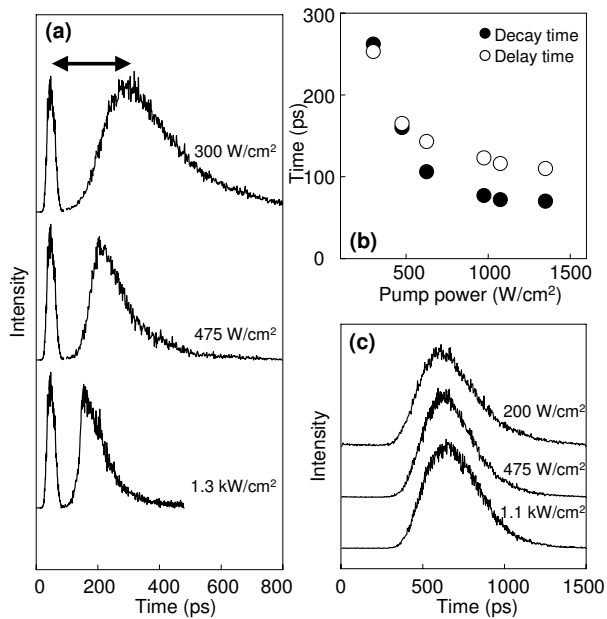}
\caption{(a) Temporal evolution of microdisk emission at 1.615 eV.
Data at different pump intensities (indicated next to each
spectrum) has been normalized and vertically shifted for clarity.
The zero of time is fixed at the arrival of the excitation laser
pulse (tuned to 1.676 eV). (b) Power dependence of decay and delay
times. Decay time, defined as the 1/eth time constant of the decay
tail, is determined from exponential fits to the data. Delay time
is defined as the interval between the peaks of the excitation and
microdisk emission pulses (arrow in (a)). Both show a marked
decrease with increasing pump power. (c) The unprocessed part of
the sample does not show any noticeable changes in either the
pulse width or the delay time as input power is increased. All
data in (a) - (c) are taken at 5 K.}
\label{fig3}
\end{figure}

Time-integrated measurements of the microdisk emission using
pulsed excitation show spectral behavior that is similar to that
obtained with CW excitation. In addition, we have measured the
temporally and spectrally resolved emission spectra at different
excitation powers (Fig. 3(a)). Both the rise and decay of the
emission become faster as the pump power exceeds the threshold
(190 W/cm$^2$ for this particular microdisk). This is shown in
Fig. 3(b) where the decay and delay time is plotted as a function
of pump power. The decay time, defined as the 1/e$^{th}$ time
constant of the decay tail, is determined from exponential fits to
the data. The delay (or rise) time is defined as the interval
between the peaks of the excitation and microdisk emission pulses
(arrow in Fig. 3(a)). The fast rise and decay dynamics is due to
the increased stimulated emission rate which is proportional to
the photon number in the cavity mode and which dominates the
emission spectra above the lasing threshold. A qualitatively
similar temporal characteristic has been reported in earlier,
spectrally-integrated dynamical studies of GaAs/(Ga,Al)As
microdisk QW lasers.\cite{luo_APL_2000} In contrast to the
shortening cavity-photon lifetime as the pump power increases, the
photon lifetime in the unprocessed IFQD sample is power
independent due to the lack of stimulated emission, as shown in
Fig. 3(c).

Finally, we comment briefly on a comparison between the behavior
of the microdisk sample with IFQDs and the control samples under
pulsed excitation. The pump-power dependence of the decay times
obtained for the control samples is almost identical to those
observed for the IFQD sample, when the pump power is scaled by the
threshold value (since the control samples have higher lasing
threshold). However, we note that there are some differences : in
the control samples the peak intensity of the lasing mode is an
order of magnitude smaller than that of the IFQD sample; the {\it
spectrally-integrated} time-resolved emission for the IFQD sample
exhibits a single peak in the time domain, whereas the control
sample with 10 nm QWs shows a second peak that appears for high
pump power after a delay of $\sim 200 - 300 $ps. The latter
behavior is similar to that reported in previous
spectrally-integrated studies of GaAs/(Ga,Al)As microdisk QW
lasers where it was ascribed to carrier diffusion to the edges of
the microdisk;\cite{luo_APL_2000}  finally, the turn-on (delay)
time for the 4.2 nm QW control sample is considerably longer than
that of the IFQD sample. As supported by  the
spectrally-integrated time-resolved data, carrier diffusion, which
is the determining factor for the delay time, is limited in the
IFQD sample compared to that seen in the control samples, which
may be attributed to the presence of the IFQDs. This limitation of carrier diffusion may possibly be responsible for the better performance (larger $Q$ and $\beta$, smaller threshold) of the microdisks with IFQDs. 

In summary, we have investigated the static and dynamic response
of a GaAs/(Ga,Al)As microdisk laser with IFQDs in the active
region. Steady-state measurements of the stimulated emission yield
a quality factor $Q \sim 5600$ and a coupling constant $\beta \sim
0.09$. The broad gain spectrum produces mode hopping between
spectrally adjacent whispering gallery modes as a function of
either temperature or excitation power. Time-resolved
photoluminescence measurements show marked differences between the
radiative decay rates in processed and unprocessed samples. The
control samples without IFQDs display modes with smaller $Q$
values and far lesser intensity. While, as mentioned earlier, the
exact effect of the IFQDs is not easy to define, it seems amply
clear that these microdisk lasers provide new templates for
exploiting the interplay between 0D states and confined photons
for quantum information processing.

This work was supported by the DARPA QuIST program, NSF-DMR and
Sun Microsystems. We thank E. Hu for critical comments, and R.
Epstein, O. Maksimov, and M. Stone for technical advice at various
stages of this project.



\begin{thebibliography}{16}
\bibitem{note1} By definition, $\beta$~ and $\xi$~ are related by a proportionality factor. We have checked that the fit yields credible estimates for the physical parameters involved in this factor.
\end{thebibliography}
\end{document}